\documentclass[11pt]{article}
\usepackage{graphicx,epsf}
\def\issue(#1,#2,#3){{\bf #1}, #2 (#3)} 

\def\opcit(#1){ {\em op. cit.}, #1}
\def\etal {\em et al.}

\def\APP(#1,#2,#3){Acta Phys.\ Polon.\ \issue(#1,#2,#3)}
\def\ARNPS(#1,#2,#3){Ann.\ Rev.\ Nucl.\ Part.\ Sci.\ \issue(#1,#2,#3)}
\def\CPC(#1,#2,#3){Comp.\ Phys.\ Comm.\ \issue(#1,#2,#3)}
\def\CIP(#1,#2,#3){Comput.\ Phys.\ \issue(#1,#2,#3)}
\def\EPJC(#1,#2,#3){Eur.\ Phys.\ J.\ C\ \issue(#1,#2,#3)}
\def\EPJD(#1,#2,#3){Eur.\ Phys.\ J. Direct\ C\ \issue(#1,#2,#3)}
\def\IEEETNS(#1,#2,#3){IEEE Trans.\ Nucl.\ Sci.\ \issue(#1,#2,#3)}
\def\IJMP(#1,#2,#3){Int.\ J.\ Mod.\ Phys. \issue(#1,#2,#3)}
\def\MPL(#1,#2,#3){Mod.\ Phys.\ Lett.\ \issue(#1,#2,#3)}
\def\NP(#1,#2,#3){Nucl.\ Phys.\ \issue(#1,#2,#3)}
\def\NIM(#1,#2,#3){Nucl.\ Instrum.\ Meth.\ \issue(#1,#2,#3)}
\def\PL(#1,#2,#3){Phys.\ Lett.\ \issue(#1,#2,#3)}
\def\PRD(#1,#2,#3){Phys.\ Rev.\ D \issue(#1,#2,#3)}
\def\PRL(#1,#2,#3){Phys.\ Rev.\ Lett.\ \issue(#1,#2,#3)}
\def\SJNP(#1,#2,#3){Sov.\ J. Nucl.\ Phys.\ \issue(#1,#2,#3)}
\def\ZPC(#1,#2,#3){Zeit.\ Phys.\ C \issue(#1,#2,#3)}

\def\bra {\langle}
\def\ket {\rangle}

\def\l {\lambda}

\def\g {\gamma}
\def\r {\rightarrow}

\def\rnot {R\!\!\!/}
\def\bar {\overline}

\def\be {\begin{equation}}
\def\ee {\end{equation}}
\def\bea {\begin{eqnarray}}
\def\eea {\end{eqnarray}}
\def\n {\nonumber}
\def\bc {\begin{center}}
\def\ec {\end{center}}

\begin{document}
\thispagestyle{empty}
\title{
\rightline  {\small{hep-ph/0302123}}
\rightline  {\small{DO-TH 03/02}}\vspace*{0.5cm}
\bf 
Simultaneous Solution to \boldmath$B\r\phi K~$\unboldmath CP Asymmetry\\
and \boldmath$B\r\eta' K, 
B\r\eta K^*~$\unboldmath Branching Ratio\\
Anomalies from R-Parity Violation}
\author{ 
{\large\bf Anirban Kundu}
\thanks{Electronic address: kundu@zylon.physik.uni-dortmund.de}
\thanks{On leave from Department of Physics, Jadavpur University, 
Kolkata 700032, India}\\
Universit\"at Dortmund, Institut f\"ur Physik, D-44221 Dortmund, Germany 
\and
{\large\bf Tandra Mitra}\\
Department of Physics, Jadavpur University, Kolkata 700032, India}

\date{\today}
\maketitle

\begin{abstract}

The branching ratios for the neutral and charged
B decay channels to $\eta' K$ and $\eta K^*$ are well above the Standard Model
expectations. Moreover, the mixing-induced CP asymmetry in $B\r\phi K_S$
is incompatible to that found in $B\r J/\psi K_S$. We investigate whether
a flavour-specific tree-level $b\r s\bar{s} s$ operator coming from
R-parity violating supersymmetry can resolve both these discrepancies, 
without jeopardizing those results which are in agreement with the
Standard Model. We found that it is possible to have a parameter space
satisfying all these requirements, including that of a low strong phase
difference compatible with the color transparency argument. 
Furthermore, we put a robust bound on the relevant
coupling, which is two orders of magnitude better than the existing one.

\end{abstract}

\noindent PACS number(s): 
11.30.Fs, 12.60.Jv, 13.25.Hw

\newpage

\section{Introduction}

The flavour structure of the Standard Model (SM) is, arguably, the best
proof that there is new physics lurking above the electroweak scale. We do
not know how this new physics should manifest itself, so one is forced to
consider all possible ways to uncover it. Moreover, any experimental
result that deviates from the SM expectation is bound to receive close
scrutiny for new physics.
Apart from the direct production of new particles, one can
look for their indirect effects, {\em i.e.}, how the operators 
at a high energy affect the low-energy observables. In this respect, the
$e^+e^-$ B factories at Cornell, SLAC, and KEK are doing a commendable job
by churning out a huge amount of data on various B decay modes, including
the branching ratios (BR) and CP asymmetries.

Even at this stage, there exist several hints that everything is not
as one should expect from the SM, if we take the experiments seriously.
The SM is in no way ruled out, since there is much scope for theoretical
uncertainty in low-energy QCD, but it will be under considerable strain
if the experimental results persist over the next few years.
In a large set of data, most of which are in perfect agreement with the SM,
there are three so-called sore thumbs sticking out: (i) The direct and
mixing-induced CP-asymmetries for the mode $B\r\pi^+\pi^-$ where Belle
and BaBar are not consistent \cite{babar-pipi,belle-pipi}, 
(ii) The abnormally high branching ratios
for the $B\r\eta'K$ and $B\r\eta K^*$
modes, incompatible with the SM prediction with
factorization \cite{pdg2002}, and 
(iii) The measured value of $\sin 2\beta$ from $B\r\phi
K_S$ \cite{babar-phik, belle-phik} \footnote{There are some other anomalies,
{\em e.g.}, lack of a consistent fit of the BRs of the $B\r PV$ channel
\cite{aleksan}. It is not clear at this moment whether this is a feature
of QCD factorization models, which was used in the analysis, or is true in
general.}. 
(Last two points are closely related; see, {\em e.g.}, \cite{soni}.)
Let us briefly comment on these anomalies. But first we wish to make our
notation clear: we use $B$ as a general shorthand for both $B^0$ and
$\bar{B^0}$, and also for the charged $B$ mesons where the mention of
charge is either unnecessary or self-evident, and 
the BRs are assumed to be averaged over the CP conjugate
states. Moreover, we use the symbol $\eta K$ to denote the
final states with either an $\eta$ or $\eta'$ and a $K$ or $K^*$ (neutral
or charged).

The first anomaly exists only for Belle --- BaBar data is completely 
consistent with the SM expectation, if one uses perturbative QCD calculations
\cite{keum-pqcd} or QCD-improved factorization model \cite{beneke}. 
However, if we take the Belle data
seriously, existence of new physics is hinted. A possible solution
is discussed in \cite{bdk2}.

The second and third anomalies are the main theme of this paper. Before
one confronts the data with any specific model, three things are
obvious. First, even if there is new physics, it is very much 
flavour-sensitive, since modes with the same topologies are not equally
affected. Secondly, the effect of new physics is of the order of unity,
considering the change in the BRs of $B\r\eta K$ modes, and the CP asymmetry
from the $B\r\phi K_S$ mode. Thus, the effective operators generated by
the new physics should better be tree-level at the high scale, unless one
prefers a strong-coupling theory. Thirdly, the new physics amplitude should
have weak (and possibly strong) phases different from the corresponding
SM amplitude to generate the observable CP asymmetries, particularly in
the $B\r\phi K_S$ channel \cite{chiang}. 

These constraints severely limit the choice of plausible new physics
options \footnote{There may be new physics with loop-induced and/or
flavour-blind operators, but their effects will be very hard to detect in the
B factories. However, some supersymmetric resolutions of the $\phi K$ puzzle
have been proposed \cite{susyphik}.}. 
Among those that survive, one of the most respectable
options is R-parity violating supersymmetry (RPV) \cite{rpvrefs1,rpvrefs2}. 
RPV automatically 
generates flavour-specific tree-level operators which can leave their 
signals in some, but not all, B decay modes. 

The effect of RPV on B decays has been extensively discussed in the literature
\cite{bdk2,cdk1,rpv-b}.
It has been proposed as a solution to the $B\r\phi K_S$
CP asymmetry puzzle \cite{alak}, 
and even a simultaneous solution to both the $\eta' K$
and $\phi K_S$ anomalies has been proposed \cite{bhaskar}.

What is new, then, in this paper? There are two main points where our
analysis differs from \cite{bhaskar}. This is the first time that one takes 
all the relevant parameters, including the final-state strong phases,
into account, and tries to explain all available data, including the
direct CP asymmetry results. Secondly, we have not looked at isolated
points in the parameter space as possible solutions, but 
have made a complete scan
over the whole parameter space, and find out the regions
which satisfy all the existing data. In fact, we 
introduce only one new physics operator, which,
we find, does the fitting quite well. Moreover, we get a stronger bound on the
relevant coupling, two orders of magnitude better than the existing one.

The discrepancies from the SM expectations show up in the BRs of different 
$B\r \eta K$ modes (see our convention before), 
and the value of $\sin(2\beta)$ extracted from
$B\r\phi K_S$ decay. The numerical values are given in Section 2.  
By itself, the CP asymmetry data has nothing wrong in it, but it is definitely
not consistent with the value of $\sin(2\beta)$ extracted from 
$B\r J/\psi K_S$ (and other charmonium channels)
even at $2\sigma$.

That the experimental numbers are not consistent with the SM predictions for
$\eta' K$ and $\eta K^*$ modes are known for a long time. Many solutions
for this are proposed, ranging from introducing a charm content of $\eta'$
or anomalous $\eta'$ coupling to gluons to the introduction of new physics like 
RPV \cite{cdk1,etaprgg}. For our calculation
we take the charm contents in both these mesons to be zero.
To have an estimate of the SM BRs for these
modes, we assume naive factorization (NF) to hold good. Note that the BRs are
more or less stable with the variation of the number of effective colors,
$N_c$, which takes into account the nonfactorizable effects \cite{ali1}. 
This, however, is not true for the generic $B\r\phi K$ modes, 
which is a penguin and the BR depends sensitively on $N_c$, falling quickly
as $N_c$ increases. Thus, if one takes a fixed value of $N_c$, the BR
predictions (within the context of the SM) for the $\eta K^*$ and $\eta' K$ 
modes are sort of reliable, but that is not the case for the $\phi K$ mode. 
In fact, it has been shown that dynamical enhancement of the penguin
operators may lead to a twofold increase of the BR over the expectation
from the NF model \cite{keum1}.
Keeping in mind such sensitivity which results from a 
numerical cancellation between different Wilson coefficients (WC), we think
it prudent to present the analysis both by including Br($B\r\phi K)$ as a 
constraint and by discarding it. We will show that the tighter constraints
on the parameter space come from the CP asymmetry data 
and the $B\r\eta K$ BRs, and relaxation of the $B\r \phi K$ BR constraint
does not affect the solutions in any perceptible way.

The paper is arranged as follows. In the next section we enlist all our
input parameters, relevant formulae and experimental data. 
In Section 3, we introduce the RPV effective
hamiltonian, and show how it affects the modes we are interested in. 
Section 4 deals with 
the numerical results, and we display the allowed regions of 
the parameter space that satisfy all data. In Section 5
we summarize and conclude.

\section{Input Parameters}

We consider only those modes which are, even in the SM, governed by the
$b\r s\bar{q} q$ transitions, with $q=u$, $d$ or $s$. The relevant effective 
four-Fermi hamiltonian reads
\be
{\cal H}_{eff} = {G_F\over\sqrt{2}}\left[ V_{ub} V_{us}^* \sum_{i=1,2} C_i
(\mu) O_i(\mu) - V_{tb} V_{ts}^*  \sum_{i=3}^{10} C_i(\mu) O_i(\mu)\right]
\ee
where the operators $O_i$ has the standard form
\bea
O_1 &=& (\bar{s}_\alpha \gamma^\mu P_L u_\beta)(\bar{u}_\beta\gamma_\mu P_L
b_\alpha),\nonumber\\
O_{3,5} &=& (\bar{s}_\alpha \gamma^\mu P_L b_\alpha)\sum_{q=u,d,s}
(\bar{q}_\beta\gamma_\mu P_{L,R} q_\beta),\nonumber\\
O_{7,9} &=& {3\over 2}(\bar{s}_\alpha \gamma^\mu P_L b_\alpha)\sum_{q=u,d,s}
(e_q\bar{q}_\beta\gamma_\mu P_{R,L} q_\beta),
\eea
with $\alpha$ and $\beta$ being the color indices, and $e_q$ the charge
of the corresponding quark.
The operators $O_{2n}$ are obtained from $O_{2n-1}$ by color-singlet 
$\leftrightarrow$ color-octet transformation. The projection operators
are $P_L(P_R) = 1-(+) \gamma_5$.   

The effective WCs for the transition $b\r s$
are evaluated at the scale $\mu = m_b/2$ at
next-to-leading-log (NLL) precision in naive dimensional regularization
(NDR) scheme, with $m_t=170$ GeV, $\alpha_s(m_Z) = 0.118$, $\alpha(m_Z)
= 1/128$, and the QCD scale $\Lambda^{(5)}_{\overline{MS}}= 225$ MeV.
They typically change by 10\% if we vary the QCD scale by about 60 MeV.
The values, which are taken from \cite{ali1}, read
\bea
&{}& C_1 = -0.33,\ C_2 = 1.16,\ C_3 = 0.022+0.003i,\ 
C_4 = -0.051-0.009i,\nonumber\\
&{}& C_5 = 0.016+0.003i, \  C_6 = -0.063-0.009i,\ C_7 
= -(1.2+1.3i)\times 10^{-4},\nonumber\\ 
&{}&C_8 = 5\times 10^{-4},\ C_9 = -(101+1.3i)\times 10^{-4},\
C_{10} = 20\times 10^{-4}.
\eea
 
At $\mu=m_b/2$, the current quark masses (in GeV) are taken to be
\be
m_u = 0.0042,\ m_d = 0.0076,\ m_s = 0.122,\ m_c = 1.5,\ m_b = 4.88.
\ee
The masses for the mesons $B^0$, $B^-$, $\pi$, $\eta$, $\eta'$, $K$ and
$K^*$ are the corresponding central values as given in
\cite{pdg2002}.

The meson decay constants (in GeV) are:
\be
f_\pi = 0.133,\ f_K = 0.158,\ f_{K^*} = 0.214,\ f_\phi = 0.233.
\ee
The $\eta$ and $\eta'$ decay constants are obtained from $f_{\eta_1} =
1.10f_\pi$ and $f_{\eta_8} = 1.34 f_\pi$ \cite{cdk1}:
\bea
f^u_\eta = {f_{\eta_8}\cos\theta\over\sqrt{6}} - 
           {f_{\eta_1}\sin\theta\over\sqrt{3}}, &{}&
f^s_\eta = - {2f_{\eta_8}\cos\theta\over\sqrt{6}} - 
           {f_{\eta_1}\sin\theta\over\sqrt{3}}, \nonumber\\ 
f^u_{\eta'} = {f_{\eta_8}\sin\theta\over\sqrt{6}} + 
           {f_{\eta_1}\cos\theta\over\sqrt{3}}, &{}&
f^s_{\eta'} = - {2f_{\eta_8}\sin\theta\over\sqrt{6}} + 
           {f_{\eta_1}\cos\theta\over\sqrt{3}}.
\eea
The mixing angle $\theta$ is taken to be $-22^\circ$. Thus the numerical
values are $f^u_\eta = 0.099$, $f^s_\eta = -0.103$, $f^u_{\eta'} = 0.051$,
$f^s_{\eta'}=0.133$. This shows why the strange quark plays a dominant
role in decays involving an $\eta'$. These values differ slightly from those
given in \cite{ali1} using a two-angle mixing scheme taking into
account the coupling of gluons to $\eta$ and $\eta'$; 
the corresponding numbers are $0.077$, $-0.112$,
$0.063$ and $0.141$ respectively. Our results do not change appreciably
if we use the latter set. 

The magnitude of the CKM elements are taken from \cite{laplace02} which
uses a fit based only on the unitarity of the mixing matrix. The error
limits are at 95\% CL.
\bea
&{}& |V_{ud}| = 0.97504\pm 0.00094,\ 
|V_{us}| = 0.2221 \pm 0.0042,\
|V_{ub}| = 0.00352 \pm 0.00103,\nonumber\\ 
&{}& |V_{cd}| = -0.2220\pm 0.0042,\ 
|V_{cs}| = 0.97422 \pm 0.00102,\
|V_{cb}| = 0.0407 \pm 0.0028,\nonumber\\ 
&{}& |V_{td}| = 0.0079\pm 0.0016,\ 
|V_{ts}| = -0.0403 \pm 0.0030,\
|V_{tb}| = 0.99917 \pm 1.2\times 10^{-4}.
\eea
We could even have used the numbers from the standard CKM fits with 
$\Delta m_B$ as one of the inputs, since, as we will show later, the
RPV couplings considered here do not affect the $B^0-\bar{B^0}$ mixing
amplitude. This, however, is not true in general; {\em e.g.}, see ref.\
\cite{bdk2}.

The transition formfactors \cite{bsw} at $q^2=0$ are given by
\be
F_0(B\r K) = 0.38;\ F_0(B\r\eta) = 0.145;\ F_0(B\r\eta') = 0.135;\
A_0(B\r K^*) = 0.32,
\ee
and $F_0(0) = F_1(0)$. One could have used the so-called `hybrid' formfactors
using both lattice QCD and light-cone QCD \cite{ali1}. However, they are
completely consistent with the numbers quoted above, and the $B\r\eta K$
BRs are only mildly affected by the latter choice. Moreover, the SM spread in
these BRs is more or less taken care of by a scan over $N_c$. 

The CP asymmetry for $B\r J/\psi K_S$ is not modified by our choice of
RPV parameters (more on this in the next section). This helps us to take
the SM value of the angle $\beta$ to be given by 
\cite{psiks}
\be
\sin(2\beta) = 0.79\pm 0.10.
\ee
We show all our results with the central value of $\sin(2\beta)$, since 
the uncertainty has negligible effect over the final results.

For the $\eta K$ modes, the data reads \cite{pdg2002,aleksan,babar0111087,
belle0207033}:
\bea
Br(B^+\r\eta'K^+) &=& (75\pm 7)\times 10^{-6}\nonumber\\
Br(B^+\r\eta {K^*}^+) &=& (25.4\pm 5.6)\times 10^{-6}\nonumber\\
Br(B^0\r\eta'K^0) &=& (58^{+14}_{-13})\times 10^{-6}\nonumber\\
Br(B^0\r\eta {K^*}^0) &=& (16.41\pm 3.21)\times 10^{-6}\nonumber\\
Br(B^+\r\eta K^+) &<& 6.9\times 10^{-6}\nonumber\\
Br(B^+\r\eta'{K^*}^+) &<& 35\times 10^{-6}\nonumber\\
Br(B^0\r\eta K^0) &<& 9.3\times 10^{-6}\nonumber\\
Br(B^0\r\eta'{K^*}^0) &<& 24\times 10^{-6}\nonumber\\
A_{CP}(B^\pm \r \eta' K^\pm) &=& 0.11 \pm 0.11 \pm 0.02 
\ ({\rm BaBar})\nonumber\\
A_{CP}(B^\pm \r \eta' K^\pm) &=& 0.015 \pm 0.070 \pm 0.009 
\ ({\rm Belle})\nonumber\\
A_{CP}^{dir}(B\r\eta' K_S) &=& -0.13\pm 0.32 ^{+0.06}_{-0.09} 
\ ({\rm Belle})\nonumber\\
A_{CP}^{mix}(B\r\eta' K_S) &=& -0.28\pm 0.55 ^{+0.07}_{-0.08} 
\ ({\rm Belle})
\eea
where 
\bea
A_{CP}(B^\pm \r \eta' K^\pm) &=& {\Gamma(B^+\r\eta' K^+)-\Gamma(B^-\r\eta' K^-)
\over \Gamma(B^+\r\eta' K^+)+\Gamma(B^-\r\eta' K^-)}\nonumber\\
A_{CP}^{dir} = {1-|\lambda|^2\over 1+|\lambda|^2},&{}&
A_{CP}^{mix} = {2Im\lambda\over 1+|\lambda|^2}
\eea
with
\be
\lambda = e^{-i\phi_M}{\bra \eta' K_S|\bar{B^0}\ket \over
\bra \eta' K_S|B^0\ket },
\ee
$\phi_M$ being the mixing phase in the $B^0-\bar{B^0}$ box ($2\beta$ in the
SM). Note that BaBar uses a convention which differs by a minus sign from 
our convention of $A_{CP}$ (see eq.\ (1) of \cite{babar0111087}), 
and the convention of Belle differs by a minus sign
too, in both direct and mixing-induced CP asymmetries. Also note 
that if $B\r\eta' K_S$ were a pure $b\r s$ penguin, the mixing-induced
CP asymmetry would have been $\sin(2\beta)$; but the presence of a tree-level
$b\r u\bar{u}s$ term makes the calculation more complicated. That is why we
have used the data from \cite{belle0207033} which gives directly the
direct and mixing-induced CP asymmetries, rather than \cite{belle-phik},
which quotes an effective value of $\sin(2\beta) = 0.76\pm 0.36$.   

Among these data, the measured BRs are definitely higher than
the theoretical predictions; the discrepancy is about a factor of 2 to 3
for the $\eta' K$ modes and almost a factor of 10 for the $\eta K^*$ modes
within the NF model (this depends on the choice of the 
regularization scale, and the effective number of colors). The expressions 
for the amplitudes can be found in the appendix, and the numerical values
of the theoretically expected BRs in tables 8 and 10, of \cite{ali1}. However,
there are other models, as the perturbative QCD \cite{keum1}, which predict
a dynamical enhancement of the penguin amplitude due to the higher-twist
corrections. Judging by the enhancement one may get in these models for
charmless modes, one expects at most to gain a factor of 2 over the
factorization amplitude. This, evidently, does not serve our purpose. Let us
just point out again that we take neither any charm content in $\eta$ or 
$\eta'$ nor any anomalous $\eta(\eta')$ coupling with gluon. 
The CP asymmetry data are consistent with the SM prediction; if
one assumes these decays to be penguin dominated as a first approximation,
the direct CP asymmetry should be zero, whereas the mixing-induced CP
asymmetry should be just $\sin(2\beta)$. The tree-level charged current
operators may change that prediction. 

The modes $B\r\phi K$ (both neutral and charged) have also been measured,
and the averaged BRs are \cite{aleksan}:
\bea
Br(B^+ \r \phi K^+) &=& (8.58\pm 1.24)\times 10^{-6},\nonumber\\
Br(B^0 \r \phi K^0) &=& (8.72\pm 1.37)\times 10^{-6},
\eea
where the near equality is expected from the isospin symmetry.
The direct CP asymmetry is measured by BaBar \cite{babar0111087} for the
charged mode and by Belle \cite{belle-phik} for the neutral mode:
\bea
A_{CP}(B^\pm\r\phi K^\pm) &=& 0.05\pm 0.20 \pm 0.03 \nonumber\\
A_{CP}(B\r\phi K_S) &=& 0.56 \pm 0.41 \pm 0.12. 
\eea
Note again that both BaBar and Belle conventions differ from ours in an
extra minus sign.
The mixing-induced CP asymmetry, which should give $\sin(2\beta)$ to a 
very good approximation, is (at $1\sigma$) \cite{babar-phik,belle-phik}:
\bea
\sin(2\beta)_{B\r\phi K_S} &=&
-0.19^{+0.52}_{-0.50}\pm 0.09 ~~{\rm (BaBar)}\nonumber\\
\sin(2\beta)_{B\r\phi K_S} &=& -0.73\pm 0.64\pm 0.18 ~~{\rm (Belle)}.
\eea
Though their central values are not compatible, one may still make an average:
\be
\sin(2\beta)_{ave} = -0.39\pm 0.41.
\ee
We use the averaged values of BR and mixing-induced CP asymmetries as
our input parameters. For the direct CP asymmetry, we use the Belle data,
but it may be observed later that the in the entire allowed parameter space 
the direct CP asymmetry is rather small (between $0.13$ and $0.25$) so
that the BaBar numbers are not in trouble. 

The BR for the $B\r\phi K$ mode is a sensitive function of $N_c$. It has
also been shown in \cite{keum1} that one needs to take into account the
annihilation and nonfactorizable contributions, which push the BR up by 
almost a factor of 2. While the BR is in perfect agreement with the
NF model, there is no model which can explain the data
on mixing-induced CP asymmetry without invoking new physics. 
For this purpose, in our analysis, 
we use the NF model to calculate the BR keeping $N_c$ a free parameter,
which we take to be the same for both $\phi K$ and $\eta K$ modes just
for simplicity. We will also comment on what happens when one relaxes the
BR constraint on $\phi K$.

\section{R-parity Violating Supersymmetry}

R-parity is a global quantum number, defined as $(-1)^{3B+L+2S}$, which
is $+1$ for all particles and $-1$ for all superparticles. In the minimal
version of supersymmetry and some of its variants, R-parity is assumed
to be conserved {\em ad hoc}, which prevents single creation or
annihilation of superparticles. However, models with broken R-parity can
be constructed naturally, and such models have a number of interesting
phenomenological consequences. 
The crucial point is
that unlike most extensions of the SM, {\em RPV contributes to B
decay amplitudes at the tree level}.  Moreover, the current bounds
\cite{rpvrefs2} on sparticle masses and couplings leave open the possiblity
that such contributions can indeed be comparable to or even larger
than the SM amplitude. It may be noted that the presence of two
interfering amplitudes of comparable magnitude is essential for a
large deviation of CP asymmetries from the SM prediction.

It is well known that in order to avoid rapid proton decay one cannot
have both  lepton number and  baryon number violating RPV model, and we 
shall work with a lepton number violating one. This leads
to slepton/sneutrino mediated B decays. Since the current lower bound on the
slepton mass is weaker than that on squark mass,
larger effects within the reach of current round of experiments are
more probable in this scenario. We start with the superpotential
\be
\label{w}
{\cal W}_{\lambda'} = \lambda'_{ijk} L_i Q_j D^c_k,
\ee
where $i, j, k = 1, 2, 3$ are quark and lepton generation indices;
$L$ and $Q$ are the $SU(2)$-doublet lepton and quark superfields and
$D^c$ is the $SU(2)$-singlet down-type quark
superfield respectively. This leads to a four-Fermi hamiltonian relevant
for B decays \cite{cdk1}:
\bea
H_{\rnot} &=&     {1\over 4} d^R_{jkn} (\bar d_n \g^\mu P_L d_j)_8
                                   (\bar d_k \g_\mu P_R b)_8\n\\
          &{ }& + {1\over 4} d^L_{jkn} (\bar d_n \g^\mu P_L b)_8
                                   (\bar d_k \g_\mu P_R d_j)_8\n\\
          &{ }& + {1\over 4} u^R_{jnk} (\bar u_n \g^\mu P_L u_j)_8
                                   (\bar d_k \g_\mu P_R b)_8
                 + {\rm H.c.}
\eea
where
\be
d^R_{jkn} = \sum_i {\l'_{ijk}{\l'}^*_{in3} \over 2m_{\tilde\nu_{Li}}^2 },
\ \ d^L_{jkn} = \sum_i {\l'_{i3k}{\l'}^*_{inj} \over 2m_{\tilde\nu_{Li}}^2 },
\ \ u^R_{jnk} = \sum_i {\l'_{ijk}{\l'}^*_{in3} \over 2m_{\tilde  e_{Li}}^2 },
\ee
and the subscript 8 indicates that the currents are in color SU(3) octet-octet
combination. 

Following the standard practice we shall assume that the RPV 
couplings are hierarchical {\em i.e.}, only one combination of the 
couplings is numerically significant.
Let us note that both the transitions $B\r\eta^{(')}K^{(*)}$ and
$B\r\phi K$ are controlled by the quark-level transitions $b\r s\bar{s}s$.
Thus, let us assume, to start with, only $d^R_{222}$ and $d^L_{222}$ to be
nonzero, as has been done in \cite{bhaskar}. Of course, $\eta K$ modes
can be fed by $b\r u\bar{u}s$ and $b\r d\bar{d}s$ transitions. Since
they affect other decay modes like $B\r\pi K$ where there is no
apparent discrepancy with SM expectations, we assume 
those operators to be vanishing.

Next let us discard $d^R_{222}$ too.  The reason
for this is that $d^R_{222}$ and $u^R_{222}$ are related by SU(2)
symmetry, and are the same if we neglect the electroweak D-term that causes
the sneutrino-slepton mass splitting (on the other hand, 
$d^R_{222}$ and $d^L_{222}$
are completely unrelated, and unless there is some underlying texture
in the RPV couplings, there is no reason why they should be equal). 
However, presence of $u^R_{222}$
generates $b\r c\bar{c}s$ transition, which in turn affects the modes
like $B\r J/\psi K_S$ which is used as a standard to extract $\sin(2\beta)$.
Since the values of $\sin(2\beta)$ extracted from different charmonium
modes, as well as from the $J/\psi \pi^0$ mode with a different quark-level
process, are almost the same \cite{belle-phik}, it is a safe assumption,
also compatible with the principle of Occam's razor, to have no RPV
contribution to that channel. Thus, the value of $\beta$ extracted
from $B\r J/\psi K_S$ can be taken to be the SM value for that angle.
The product coupling $d^L_{222}$ does not contribute to the $B^0-\bar
{B^0}$ box, so that there is no scope to have an extra box amplitude, in
contrast to the situation, {\em e.g.}, in \cite{bdk2} (but
$d^L_{222}$ contributes to the $B_s$ box; this is discussed later). 
The QCD corrections are easy to implement:
the short-distance QCD corrections
enhance the $(S-P)\times(S+P)$ RPV operator by approximately
a factor of 2 while running from the slepton mass scale 
(assumed to be at 100 GeV) to $m_b$ \cite{bagger}. 

The RPV amplitude for $B\r\eta' K$ is given by
\be
M_{\eta' K}^{\rnot} = {1\over 4} d^L_{222} \left[R_1\left(A_{\eta' K}^s
-A_{\eta' K}^u\right) - {1\over N_c}A_{\eta' K}^s\right]
\ee
and that for $B\r\eta K^*$ is 
\be
M_{\eta K^*}^{\rnot} = {1\over 4} d^L_{222} \left[R_2\left (A_{\eta K^*}^s
-A_{\eta K^*}^u\right) - {1\over N_c}A_{\eta K^*}^s\right],
\ee
where
\bea
R_1 &=& {m_{\eta'}^2\over m_s (m_b-m_s)}\nonumber\\
R_2 &=& -{m_{\eta}^2\over m_s (m_b+m_s)}
\eea
and
\bea
A_{\eta' K}^{u,s} &=& f_{\eta'}^{u,s} F_0^{B\r K}(m_{\eta'}^2)(m_B^2-m_K^2)
\nonumber\\
A_{\eta K^*}^{u,s} &=& 2 f_{\eta}^{u,s} m_{K^*} A_0^{B\r K^*}(m_\eta^2)
(\epsilon_{K^*}.p_{\eta}).
\eea
For $B\r\phi K$ (both neutral and charged channels), the RPV amplitude
is
\be
M_{\phi K}^{\rnot} = {1\over 4N_c} d^L_{222} A_{\phi K}
\ee
with
\be
A_{\phi K} = 2 f_\phi m_\phi F_0^{B\r K}(m_\phi ^2) (\epsilon_{\phi}.p_K) 
\ee
All these amplitudes are calculated at the slepton mass scale, and, as
stated earlier, should be multiplied roughly by a factor of 2 when we compute
their effects at $m_b$.

The product coupling $d^L_{222}$ can in general be complex, which we write 
as
\be
d^L_{222} = |d^L_{222}| exp(i\phi_{\rnot}).
\ee
In our analysis we vary this phase over the range 0 to $\pi$, and include the
effects of $\pi \leq \phi_\rnot \leq 2\pi$ by allowing $|d^L_{222}|$ to take
both positive and negative values. 

For generic $B\r\eta K$ modes, even the SM has two factorizable amplitudes,
tree and penguin. Following the color transparency argument which predicts
the strong phase difference $\Delta\delta_{SM}$ 
between them to be small \cite{beneke,bjorken}, 
we take $\Delta\delta_{SM} = 0$. We vary the
strong phase difference $\Delta\delta$ 
between the SM and the RPV amplitudes over a range
of 0 to $2\pi$, but expect to find solutions allowing $\Delta\delta$ 
to be near 0 or $2\pi$, which is theoretically pleasing. 
Note that if all strong
phase differences are exactly zero, there should not be any direct CP
violation; indeed, with color transparency expectations, we should get a 
small $A_{CP}^{dir}$, which anyway is perfectly allowed by data. For 
simplicity we take $\Delta\delta$ to be the same for all
channels to be considered.

\section{The Analysis}

Our input parameters are specified in Sections 2 and 3. We scan 
the CKM element $V_{ub}$, which has an almost $30\%$ uncertainty, over
its entire range. We also vary $1/N_c$ from 0.1 to 1. The weak phase 
$\phi_\rnot$
associated with $d^L_{222}$ is scanned over 0 to $\pi$ and the strong
phase difference $\Delta\delta$
between SM (tree or penguin) and RPV over 0 to $2\pi$ (we, however,
present our results for the range $-\pi/6 < \Delta\delta < \pi/6$, motivated
by the color-transparency argument). 
The CKM angle $\gamma$ is varied between 0 and $\pi-\beta$ where
$\beta=0.5\arcsin(A_{CP}(B\r J/\psi K_S))$. 

\begin{figure}[htbp]
\vspace{-10pt}
\centerline{\hspace{-3.3mm}
\rotatebox{-90}{\epsfxsize=8cm\epsfbox{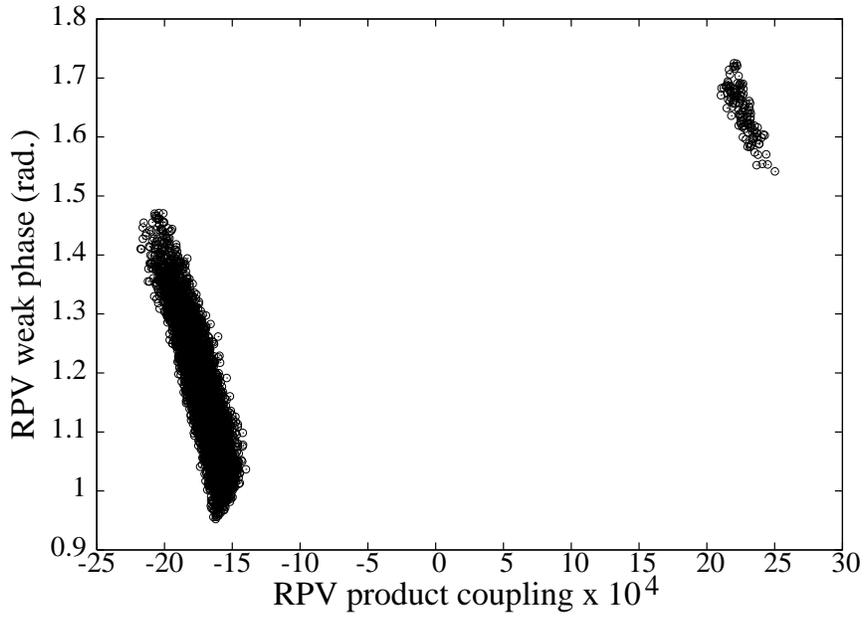}}}
\hspace{3.3cm}\caption[]{\small The allowed parameter space for
$\lambda'_{i32}\lambda'_{i22}$ and $\phi_\rnot$.
For more details, see text.}  \protect\label{fig1}
\end{figure}
\begin{figure}[htbp]
\vspace{-10pt}
\centerline{\hspace{-3.3mm}
\rotatebox{-90}{\epsfxsize=8cm\epsfbox{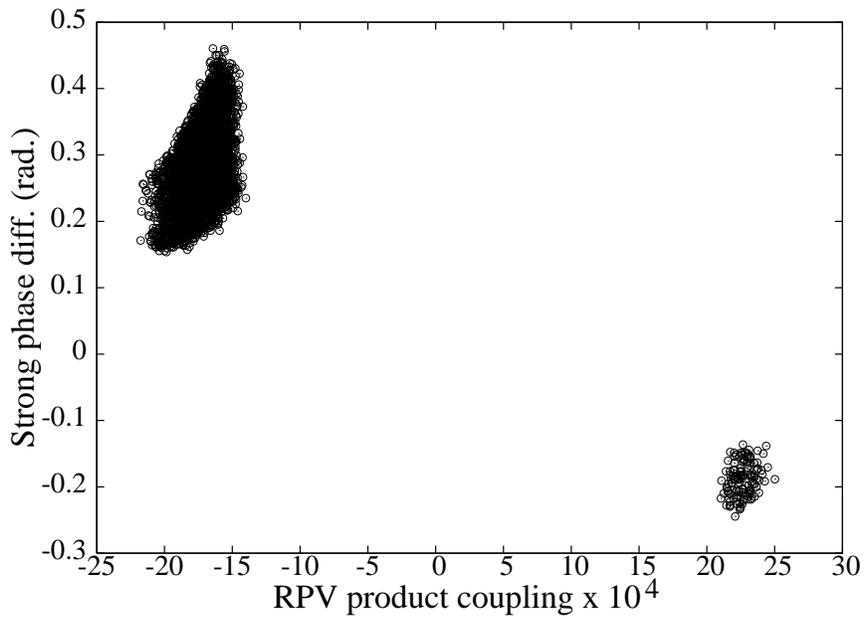}}}
\hspace{3.3cm}\caption[]{\small One can indeed have solutions with
small difference in strong phases of the SM and the RPV amplitudes.}
\protect\label{fig2}
\end{figure}

The following constraints were applied: (i) BR
for the modes $\eta' K^+$, $\eta' K^0$, $\eta {K^*}^+$,
$\eta {K^*}^0$ and $\phi K_S$, (ii)
the direct CP asymmetry for $B^\pm \r \eta' K^\pm$ from BaBar, (iii)
the direct CP asymmetry for $B \r \phi K_S$ from Belle, and
(iv) the average value of BaBar and Belle for
$\sin(2\beta)$ extracted from $B\r\phi K_S$. Constraints (ii) and (iii)
are applied with the rationale that they have larger error bars and we wish
to check whether the data from the other experiment with smaller errors
can be accomodated. The direct and mixing-induced CP asymmetries for
$B\r\eta' K_S$ are not imposed as constraints but one can easily check
from the figures that most of the allowed region is perfectly compatible 
with the data.

Our results are shown in figures 1-4. Let us note the salient features
of the analysis.

\begin{figure}[htbp]
\vspace{-10pt}
\centerline{\hspace{-3.3mm}
\rotatebox{-90}{\epsfxsize=8cm\epsfbox{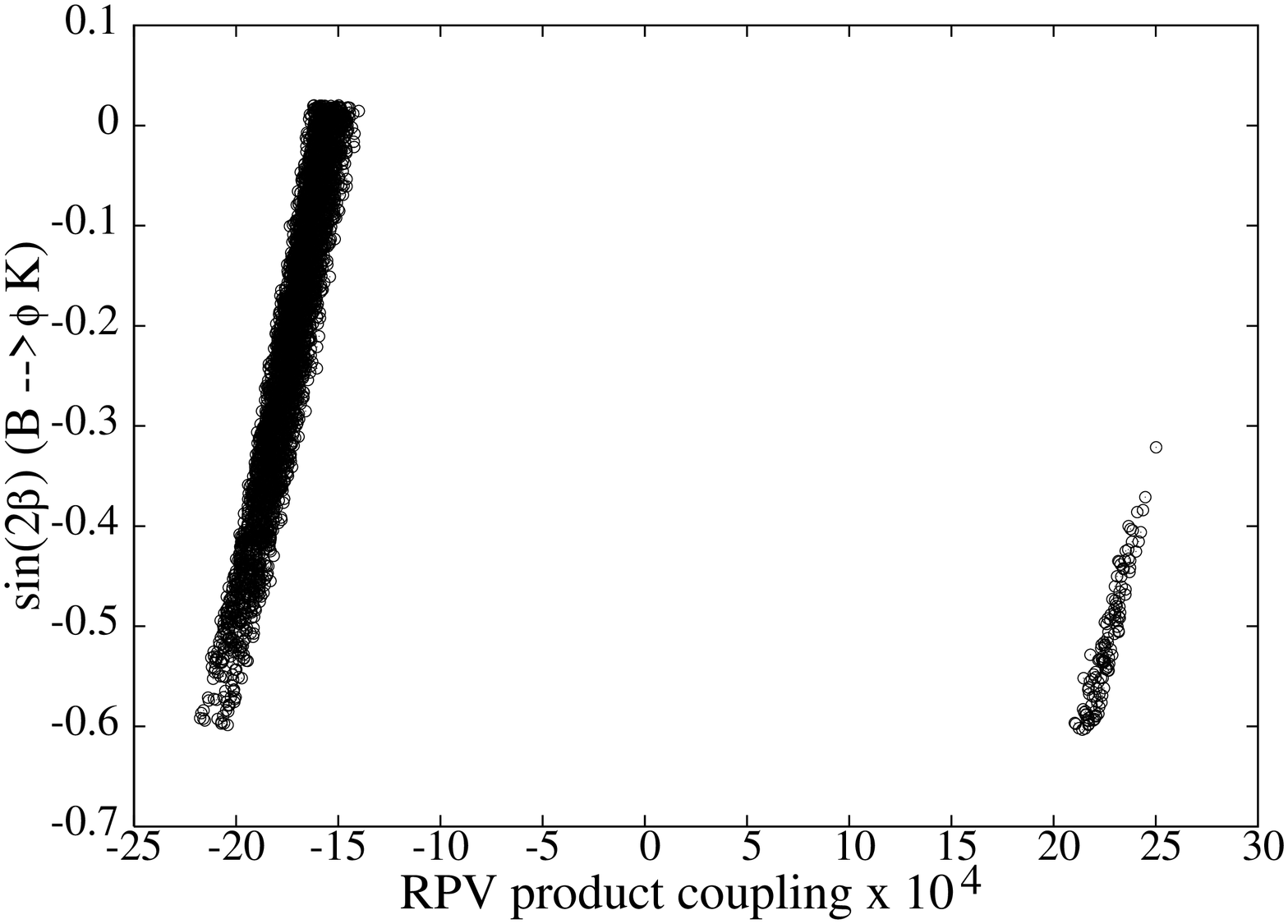}}}
\hspace{3.3cm}\caption[]{\small The range of $\sin(2\beta)$ as extracted
from the $B\r\phi K_S$ decay.}
\protect\label{fig3}
\end{figure}
\begin{figure}[htbp]
\vspace{-10pt}
\centerline{\hspace{-3.3mm}
\rotatebox{-90}{\epsfxsize=8cm\epsfbox{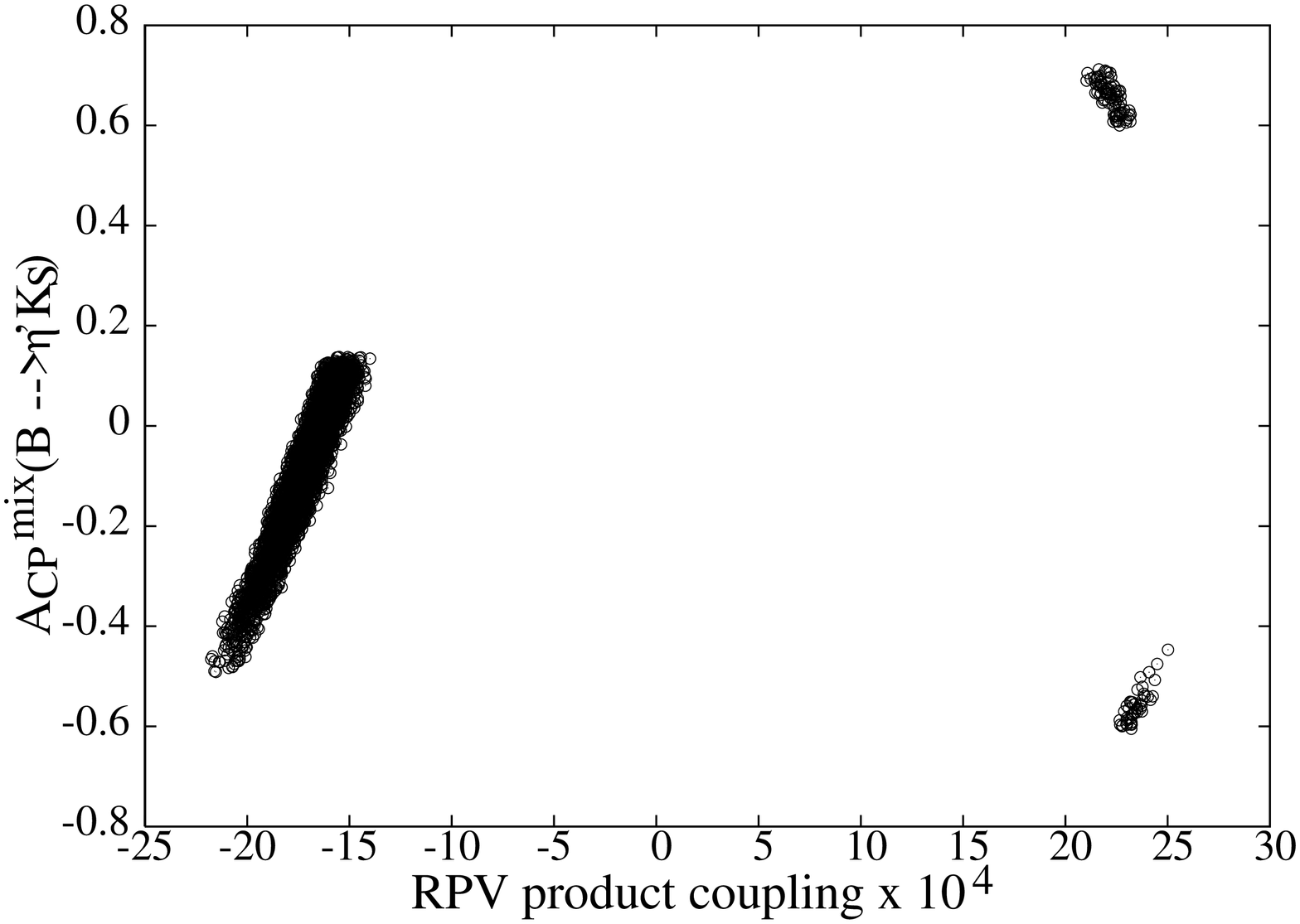}}}
\hspace{3.3cm}\caption[]{\small The mixing-induced CP asymmetry for
$B\r\eta' K_S$. Note that upper half of Band II is disallowed from the
Belle data.}
\protect\label{fig4}
\end{figure}

\begin{itemize}
\item It is known that the nonfactorizable effects in these decays can
be substantial. To account for that, we have taken $N_c$ as a free
parameter, and not stuck to its QCD value of 3. However, the Wilson 
coefficients are evaluated with $N_c = 3$. 
It appears that there is a significant nonfactorizable contribution in
$B\r\phi K$, since we have obtained the fit only for $0.15\leq 1/N_c 
\leq 0.25$. Note that $\eta K$ channels are $N_c$ stable, so the constraint
only comes from the $\phi K$ mode. This is in conformity with the analysis
in \cite{keum1}. 

\item There are two possible bands of solutions, as can be seen from
Fig.\ 1. The left-hand side band, with more points and more width, is for 
negative values of $d^L_{222}$ and $\phi_\rnot$ in the first quadrant (this we 
will call Band I). There is a second narrow band (Band II) for positive
values of $d^L_{222}$ and $\phi_\rnot$ in the second quadrant. This shape
is essentially controlled by the BRs and CP asymetries
of different $\eta K$ channels. 

 We found a much stronger constraint on the product coupling
$|\lambda'_{i32} {\l'}^*_{i22}|$:
\be
|\lambda'_{i32} {\l'}^*_{i22}| \leq 2.3\times 10^{-3}.
\ee
If we take all these experimental data seriously, it is possible to get
even an upper bound:
\be
|\lambda'_{i32} {\l'}^*_{i22}| \geq 1.3\times 10^{-3}.
\ee
The present bound, as quoted in \cite{allanach} for the third slepton
generation, is only $0.23$, and that too assuming squarks at 100 GeV. 
If one has 300 GeV squarks, the bound gets weaker by a factor of 9. However,
one gets a comparatively better constraint from $B_s-\bar{B_s}$ mixing,
since nonzero values of $\lambda'_{i32}$ and $\lambda'_{i22}$ can generate
a second amplitude for the box, with two sneutrinos and two right-handed
strange quarks flowing inside the box. Taking $\Delta m_{B_s}$ to be completely
saturated by the RPV contribution, and using the experimental lower limit, one
gets a bound on the product coupling which is approximately $1.5\times
10^{-2}$. The main loophole in the analysis is the fact that
only an experimental upper bound can generate an upper
bound on some unknown
parameter; moreover, it is questionable to neglect the SM contribution
completely. Even then we have an improvement by an order of magnitude.  

\item Figure 2 shows that $\Delta\delta$ should be positive for Band I
and negative for Band II. It can be checked easily that this ensures a 
positive direct CP asymmetry in all channels, and a negative mixing-induced
CP asymmetry in $B\r\phi K_S$. Figure 3 shows the range of $\sin(2\beta)$
in this model, which is between the upper limit of $0.02$ and $-0.6$. 
The central value of Belle, however, cannot be reproduced.  

\item Figure 4 shows the mixing-induced CP asymmetry for $B\r\eta' K_S$. 
Note that imposition of the Belle data means a significant portion (the 
region in the upper right-hand corner) of
the already weak Band II is ruled out, while Band I is completely allowed.
Still, one cannot rule out Band II completely.

\item The CP asymmetry in $B^\pm\r\eta' K^\pm$ is found to lie between
$0.115$ and $0.22$ (the upper limit) for Band I, and between $0.14$ and 
$0.22$ for Band II. The direct CP asymmetry in the neutral channel
lies between $0.13$ to $0.24$. This is definitely compatible with the
Belle data, but again the central value is far away. One must wait for the
error bars to come down.

\item The direct CP asymmetry for $B\r\phi K_S$ lies between $0.13$ and
$0.25$ for Band I and between $0.13$ and $0.22$ for Band II. This is in
perfect harmony with the BaBar data too. 

Let us try to understand why we get a nonzero $A_{CP}^{dir}$. This is due
to the constraint put by $A_{CP}^{mix} (B\r\phi K)$. To see this qualitatively,
let us assume that there is only one SM amplitude and
both weak and strong phase differences vanish so that there is no direct
CP asymmetry. It is easy to see that in that case there is no change in
the prediction for $\sin(2\beta)$, which is given only by the phase in the
$B^0-\bar{B^0}$ box.

\item The angle $\gamma$ can take any value upto $\pi-\beta$ for Band I.
On the other hand, it can be only in the second quadrant for Band II. The
origin of such a pattern can easily be traced back to the expressions of
BRs and CP asymmetries of the $\eta K$ modes. Ref.\ \cite{laplace02}
estimates $40^\circ < \gamma < 78^\circ$ at 95\% CL from different $B\r\pi\pi$
and $B\r K\pi$ modes. These modes are not affected by $d^L_{222}$, so one 
concludes that if there are no other nonzero RPV couplings,
Band II is completely ruled out. Thus, if the value of $\sin(2\beta)$
extracted from $B\r\eta' K_S$ converges towards that found from 
$B\r J/\psi K_S$, this solution will be in trouble. However, the error bars
are too large to draw any definite conclusion right now.

\item As a check, we redo the analysis switching the $B\r\phi K$ BR constraint
off. There is no substantial change in the result, except that the lower
bound on $A_{CP}(B^\pm\r\eta' K^\pm)$ marginally decreases to $0.11$. 
Thus we consider our result to be fairly robust with respect to hadronic
uncertainties. However, there is no upper limit on $1/N_c$ anymore.

\end{itemize}  

We have also explicitly checked that BRs of the so far unobserved 
$\eta K$ channels remain below their experimental upper
bounds.  

\section{Summary and Conclusion}

We found that a minimal set of RPV couplings, compatible with all present
data, can explain the BR anomalies for $B\r\eta' K$ and $B\r\eta K^*$,
and the unusual value of $\sin(2\beta)$ from $B\r\phi K_S$. This is mainly
due to the fact that RPV contributes to B decays at tree-level. A nice
feature is that one can have points where the strong phase difference
between RPV and SM is small, which is what one expects from the 
color transparency argument. 

With more data pouring in, one can significantly shrink the allowed
parameter space for RPV. However, even at present we get sufficiently
strong bound on the relevant product coupling --- better by two orders
of magnitude at least. 

What other effects are mediated by the product coupling $\lambda'_{i32}
\lambda'_{i22}$? This generates a top decay channel $t\r c\bar{s}s$, whose
strength is, unfortunately, only a few per cent of the corresponding SM
channel. Thus, the signal is essentially unobservable. 
A better signal may come from the strange squark mediated semileptonic
top decays $t\r c\ell^+\ell^-$ ($\ell = e,\mu,\tau$).
This also generates
$b\r s\nu\bar{\nu}$ or $B_s\r \nu\bar{\nu}$ decays. The third effect is
a new box amplitude in $B_s-\bar{B_s}$ mixing. Here, also, the amplitude is
only at a few per cent level compared to the SM amplitude, so we do not
expect any significant CP asymmetry in channels like $B_s\r J/\psi\phi$
even if $\phi_\rnot$ is large. 

We have not discussed the product coupling $d^R_{222}$ which can also explain
the anomalies that are studied in this paper. As we mentioned before,
that coupling also generates the SU(2) conjugate transition $b\r c\bar{c} s$.
This may jeopardize the predictions from the $B\r J/\psi K_S$ channel, for
example. However, a nonzero phase in $d^R_{222}$ should show up in the
$B_s$ box, {\em i.e.}, one may get a significant CP-asymmetry in the
channel $B_s\r J/\psi\phi$, contrary to the SM expectation. 
The reason is simple: RPV contributes to both the box amplitude and to
the decay $b\r c\bar{c} s$. The rare decays $b\r s\ell^+\ell^-$ or $B_s\r
\ell^+\ell^-$ also receive a tree-level contribution 
from $d^R_{222}$ and may be pushed up to the observable level.  

At present, one has to wait for the errors to come down.
This may rule out part or
all of one or both bands (existence of Band II is already under threat from
the fit of $\gamma$, as we have shown). However, if the bands still 
remain allowed, one 
should look for any unexpected CP asymmetry signal in $B_s$ decays. Only
such correlated studies can unravel the exact nature of new physics.

\vspace*{1cm}

\centerline{\bf Acknowledgement}

We thank S. Oh for his comments on the manuscript.
AK has been supported by the BRNS grant 2000/37/10/BRNS of DAE,
Govt.\ of India, by 
the grant F.10-14/2001 (SR-I) of UGC, India, and
by the fellowship of the Alexander von Humboldt Foundation.


\begin{thebibliography}{99}
\bibitem{babar-pipi}
 B. Aubert {\etal} (BaBar Collaboration), 
\PRL(89,281802,2002).

\bibitem{belle-pipi}
K. Abe  {\etal} (Belle Collaboration), hep-ex/0301032.

\bibitem{pdg2002}
K. Hagiwara {\etal}  (Particle Data Group Collaboration),
\PRD(66,010001,2002).

\bibitem{babar-phik}
B. Aubert {\etal} (BaBar Collaboration),
hep-ex/0207070.

\bibitem{belle-phik}
K. Abe  {\etal} (Belle Collaboration), hep-ex/0207098.

\bibitem{aleksan}
R. Aleksan {\etal}, hep-ph/0301165.

\bibitem{soni}
D. London and A. Soni, \PL(B407,61,1997);
D. Atwood and A. Soni, \PRL(79,5206,1997).

\bibitem{keum-pqcd}
G.P. Lepage and S. Brodsky, \PRD(22,2157,1980);
J. Botts and G. Sterman, \NP(B325,62,1989); 
Y.Y. Keum, H.n. Li and A.I. Sanda, \PL(B504,6,2001), 
\PRD(63,054008,2001);
Y.Y. Keum and A.I. Sanda, hep-ph/0209014.


\bibitem{beneke}
M. Beneke {\etal}, \PRL(83,1914,1999), 
\NP(B606,245,2001).

\bibitem{bdk2}
G. Bhattacharyya, A. Datta and A. Kundu, hep-ph/0212059.

\bibitem{chiang}
C.W. Chiang and J.L. Rosner, hep-ph/0302094.

\bibitem{susyphik}
E. Lunghi and D. Wyler, \PL(B521,320,2001);
M.B. Causse, hep-ph/0207070;
G. Hiller, \PRD(66,071502,2002);
M. Ciuchini and L. Silvestrini, \PRL(89,231802,2002);
M. Raidal, \PRL(89,231803,2002);
J.P. Lee and K.Y. Lee, hep.ph/0209290;
S. Khalil and E. Kou, hep-ph/0212023;
S. Baek, hep-ph/0301269.


\bibitem{rpvrefs1}
G. Farrar and P. Fayet, \PL(B76,575,1978); 
S. Weinberg, \PRD(26,287,1982); 
N. Sakai and T. Yanagida, \NP(B197,533,1982); 
C. Aulakh and R. Mohapatra, \PL(B119,136,1982).

\bibitem{rpvrefs2} For recent reviews, see G. Bhattacharyya, Nucl.
  Phys. (Proc. Suppl.) {\bf B52A}, 83 (1997); hep-ph/9709395; 
H. Dreiner, hep-ph/9707435; 
R. Barbier {\etal}, hep-ph/9810232.

\bibitem{cdk1}
D. Choudhury, B. Dutta and A. Kundu, \PL(B456,185,1999).

\bibitem{rpv-b}
G. Bhattacharyya and A. Raychaudhuri, \PRD(57,3837,1998);
D. Guetta, \PRD(58,116008,1998); 
G. Bhattacharyya and A. Datta, \PRL(83,2300,1999); 
G. Bhattacharyya, A. Datta and A. Kundu, \PL(B514,47,2001); 
D. Chakraverty and D. Choudhury, \PRD(63,075009,2001);
D. Chakraverty and D. Choudhury, \PRD(63,112002,2001);
J.P. Saha and A. Kundu, \PRD(66,054021,2002).

\bibitem{alak} A. Datta, 
\PRD(66,071702,2002).

\bibitem{bhaskar}
B. Dutta, C.S. Kim and S. Oh,
\PRL(90,011801,2003).

\bibitem{etaprgg}
D. Atwood and A. Soni, \PL(B405,150,1997);
A. Ali {\etal}, \PL(B424,161,1998);
M. Beneke and M. Neubert, \NP(B651,225,2003).

\bibitem{ali1} A. Ali, G. Kramer and C.-D. L\"{u},  \PRD(58,094009,1998).

\bibitem{keum1}
C.H. Chen, Y.Y. Keum and H.n. Li,
\PRD(64,112002,2001).

\bibitem{laplace02}
S. Laplace, hep-ph/0209188.

\bibitem{bsw}
M. Wirbel, B. Stech and M. Bauer,
\ZPC(29,637,1985).

\bibitem{psiks}
B. Aubert {\etal} (BaBar Collaboration), 
\PRL(89,201802,2002);
K. Abe {\etal} (Belle Collaboration), \PRD(66,032007,2002).

\bibitem{babar0111087}
B. Aubert {\etal} (BaBar Collaboration), \PRD(65,051101,2002).

\bibitem{belle0207033}
K.F. Chen {\etal} (Belle Collaboration), \PL(B546,196,2002).



\bibitem{bagger}
J.L. Bagger, K.T. Matchev and R.J. Zhang, \PL(B412,77,1997);
M. Ciuchini {\etal}, \NP(B523,501,1998).

\bibitem{bjorken} J.D. Bjorken, Nucl. Phys. (Proc. Suppl.) {\bf B11},
325 (1989).

\bibitem{allanach}
B.C. Allanach, A. Dedes and H.K. Dreiner,
\PRD(60,075014,1999).



\end{thebibliography}
\end{document}